\documentstyle[prb,aps,multicol,epsf]{revtex}
\begin{document}
\title{Duality Relation among Periodic Potential Problems in
the Lowest Landau Level}
\draft
\author{K. Ishikawa, N. Maeda, T. Ochiai, and H. Suzuki}
\address{Department of Physics, Hokkaido University, Sapporo 060, Japan}
\maketitle
\begin{abstract}
Using a momentum representation of  a magnetic 
von Neumann lattice,
we study a two-dimensional electron in a uniform magnetic field
and obtain one-particle spectra of various periodic short-range potential
problems in the lowest Landau level.
We find that the energy spectra  satisfy
a duality relation between a period of the potential and a magnetic
length.
The energy spectra consist of the Hofstadter-type bands and flat bands.
We also study the connection between a periodic short-range potential
problem and a tight-binding model.
\end{abstract}
\pacs{71.25.-s, 73.20.Dx}

A physical system sometimes shows a peculiar fractal structure
of the spectrum if the system has two scales of periods.
In the nearest-neighbor (NN) tight-binding model with a magnetic
flux, Hofstadter obtained one-particle spectra of butterfly-shape
and discovered a multi-fractal structure.\cite{a}
Experimentally, it may be a challenging theme to observe the
spectrum of the fractal structure.
Actually, interesting phenomena have been observed in experiments of
lateral super lattices.\cite{b,c,ca}
The system with the cosine potential has been studied well
theoretically.\cite{d}
In the case of the antidot array, however, the potential term
includes a large number of the cosine functions with different wave
lengths.\cite{fgk}
The steeper the potential of the antidot is, the harder to solve the 
eigenvalue problem is.
Silberbauer\cite{silb} and K\"uhn {\it et al}.\cite{kuhn}
studied a steep antidots potential problem. 
Huang {\it et al}. \cite{hgm} solved the periodic short-range potential 
problem in a finite system numerically. Analytic properties of the problem 
are unknown.

In the present paper, we study eigenvalue problems of the two-dimensional
electron systems  defined on an infinite plane 
with the periodic short-range potentials in the
lowest Landau level (LLL) analytically and numerically.
We find Hofstadter-type bands and flat bands.
The former bands satisfy a duality relation.
Namely, the spectrum at one value of $t$ is connected with that of $1/t$,
where $t$ is the flux $\Phi$ penetrating the unit cell of the periodic 
potential normalized by the unit flux $\Phi_0$, {\it i.e.} $t=\Phi/\Phi_0$.
Our duality is concerned with flux and is different from the Aubry and 
Andre's duality,\cite{ahk} which is concerned with  hopping strengths 
in the NN tight-binding model.
We also study the connection between a periodic short-range potential
problem and a tight-binding model.

The magnetic von Neumann lattice\cite{e,ha} is a representation of the
quantum system in a uniform magnetic field and has  quite useful
properties in studying the system with  periodic potentials.
In particular, its lattice structure varies and 
we can select the suitable one in accordance with 
the potential lattice.
Using a momentum representation of the von Neumann lattice,\cite{i} 
we give a proof of the duality and show 
energy spectra in the periodic potential problems.

In a two-dimensional system under a uniform magnetic field $B$,
 the von Neumann lattice basis 
 is formed by the  direct product of harmonic 
oscillator eigenfunctions
$\vert f_l\rangle$ of relative coordinates ($\xi,\eta$) and
coherent states $\vert \alpha_{mn} \rangle$ of guiding center coordinates
($X,Y$).
The coherent states are defined by
\begin{equation}
(X+iY)\vert\alpha_{mn}\rangle=az_{mn}\vert\alpha_{mn}\rangle,
\end{equation}
where $a=\sqrt{2\pi\hbar/eB}$, $z_{mn}=m\omega_x+n\omega_y$ 
 ($m,n \in {\bf Z}$) and
$\omega_x ,\omega_y$ are complex numbers that satisfy
${\rm Im}[\omega_x^*\omega_y]=1$.\cite{pb}
For our purpose, it is convenient to use the Fourier transformed basis
as
\begin{equation}
\vert l,{\bf p}\rangle=\frac{1}{\beta({\bf p})}\sum_{m,n}e^{ip_xm+ip_yn}
\vert f_l \otimes \alpha_{mn}\rangle,
\end{equation}
where $\beta({\bf p})$ is a normalization factor defined by
\begin{equation}
\beta({\bf p})=\left(2{\rm Im}\tau\right)^{1\over4}
e^{i{\tau\over4\pi}p_y^2}
\vartheta_1({p_x+\tau p_y\over2\pi}\vert\tau),
\end{equation}
and $\tau=-\omega_x / \omega_y$.
For $\tau=i$, the von Neumann lattice becomes a square lattice. 
For $\tau=e^{i2\pi /3}$, it becomes a triangular lattice.
The Fourier transformed basis is an orthonormal set of extended states  
and obeys a nontrivial boundary condition 
\begin{equation}
\vert l,{\bf p}+2\pi {\bf N}\rangle=e^{-i\phi (p,N)}
\vert l,{\bf p}\rangle ,
\end{equation}
where $\phi (p,N)=\pi (N_x+N_y)-N_y p_x$. 


If the magnetic von Neumann lattice has a periodicity
commensurate with the periodicity of the external potential $V({\bf x})$, 
the one-body potential problem becomes easy to treat.
This happens when $t$ is equal to $q/p$
with relatively prime integers  $p,q$.
Let us consider an arbitrary regular lattice of short-range
potentials of $t=q/p$:
\begin{equation}
V({\bf x})=a^2V_0\sum_N\delta^{(2)}(z+aN_x q \omega_x+a {N_y\over p} \omega_y).
\label{peri}
\end{equation}
The matrix element of the potential in the LLL is given by 
\begin{eqnarray}
& &\langle 0,{\bf p}\vert V({\bf x})\vert 0,{\bf p'}\rangle  \\
&=&\frac{V_0}{q} \sum_{r_p,r_q,{\bf N}} \beta (p_x-2\pi{r_p\over p},p_y) 
\beta^* (p_x-2\pi({r_p\over p}+{r_q\over q}),p_y)  \nonumber \\
&\times &(2\pi)^2 
\delta (p_x' -p_x+2\pi({r_q\over q}+ N_x))
\delta (p_y' -p_y +2\pi N_y) e^{i \phi(p',N)}, \nonumber
\end{eqnarray}
where $r_p=0,1,...,p-1$ and  $r_q=0,1,...,q-1$.
Here, we study the eigenvalue equations in the LLL.
The eigenvalue equation becomes 
\begin{equation} 
D^\dagger D\psi=\epsilon \psi,
\label{eig1}
\end{equation}
where a $p\times q$ matrix $D$ is defined by 
\begin{equation}
(D({\bf p}))_{r_p r_q}=\beta^*(p_x-2\pi({r_p\over p}+{r_q\over q}),p_y),
\label{dd}
\end{equation}
and  $D^\dagger D$ is a $q\times q$ Hermitian matrix.
The magnetic Brillouin zone (MBZ) 
is the region where $\vert p_x\vert<\pi/q$ and $\vert p_y\vert<\pi$.
Each band has a $p$-fold degeneracy. 
Consequently, the fundamental region of the MBZ is the region where 
$\vert p_x\vert<\pi/pq$ and $\vert p_y\vert<\pi$.

We should note that the rank of $D^\dagger D$ is min($p,q$) generally.
If $p<q$,  the band splits into $p$ subbands and 
one flat band which  corresponds to the  zero modes of $D$. 
The number of the zero modes is $q-p$.
By a linear transformation, $\psi'=D\psi$, 
the eigenvalue equation (\ref{eig1}) becomes
\begin{equation}
DD^\dagger\psi'=\epsilon\psi'.
\label{eig2}
\end{equation}
$DD^\dagger$ is a $p\times p$ Hermitian matrix.
Equation (\ref{eig1}) is equivalent to Eq.~(\ref{eig2}) except for
the zero modes.
Apparently, Eq.~(\ref{eig2}) is obtained by
taking the complex conjugate and
interchanging $p$ and $q$ of Eq.~(\ref{eig1}).
Therefore, there exists a duality relation between two problems of
$t=q/p>1$ and $t=p/q<1$. Actually, the energy spectra $E$ obey the
relation
\begin{equation}
E(t)={1\over t}E({1\over t}),
\label{dual}
\end{equation}
except for the flat bands.
If $p>q$, Eq.~(\ref{dual}) is also obtained.
Thus, the duality relation is proved in an arbitrary regular lattice of 
short-range potentials. 
The self-dual point is $t=1$ and the physical quantity has
critical behavior near the point.

Before solving  the eigenvalue problems numerically,
we clarify the connection between a periodic 
short-range potential problem and a tight-binding model in the LLL.
Let us suppose a square lattice potential 
$V({\bf x})=V_0 {\rm exp}(2\pi i(mx+ny)/b)$, where 
$t=(b/a)^2=q/p$.
The potential energy term in the second quantized form  becomes
\begin{eqnarray}
&V_0&\int_{\rm MBZ}{d^2p\over(2\pi)^2}
\sum_{r_q}c^\dagger_{r_q}({\bf p})c_{r_q+m}({\bf p})
\nonumber\\
&\times&e^{-{\pi\over2t}(m^2+n^2)+in(p_x-{2\pi\over t}
r_q)-i{m\over t}(p_y+\pi n)}
\label{potpot}
\end{eqnarray}
where
\begin{equation}
c_{r_q}({\bf p})=b_0(p_x+\pi
-2\pi{p\over q}r_q,p_y)e^{-i\pi{p\over q}r_q}
\end{equation}
and $b_l({\bf p})$ is the annihilation operator of the state 
$\vert l,{\bf p}\rangle$. 
Equation (\ref{potpot}) is equivalent to
the tight-binding Hamiltonian with the following hopping term:
\begin{equation}
V_{\rm hop}=
V_0e^{-{\pi\over2t}((m_2-m_1)^2+(
n_2-n_1)^2)+i{\pi\over t}(m_2+m_1)(n_2-n_1)},
\end{equation}
where $m_2-m_1=m$ and $n_2-n_1=n$.
This hopping term has a flux  per  unit cell
of a square lattice $2\pi/t$ and the hopping strength varies with $t$.
If  $(m,n)=(\pm1,0)$, $(0,\pm1)$, Eq.~(\ref{potpot})  becomes
the NN tight-binding Hamiltonian.
In a square lattice of short-range potentials,
the summation is taken over all $(m,n)$ with an equal weight.
Therefore,
the result becomes a tight-binding Hamiltonian with finite-range
hopping terms
\begin{equation}
{1\over t}\sum_{m,n}c^\dagger(m_2,n_2)V_{\rm hop}(m_2,n_2;m_1,n_1)
c(m_1,n_1).
\end{equation}
The hopping range is about $\sqrt{t}$ and the number of relevant terms
increases linearly with  $t$.


In a regular lattice,
the one-particle spectrum is given by the
solution of Eq.~(\ref{eig1}).
Since the matrix element of  $D^\dagger D$ is given in the analytic form,
it is easy to solve numerically.
Figure 1 shows the spectra for the square and triangular lattices.
The points at $E=0$ correspond to the original LLL.
There are two marked structures in these figures : 
Hofstadter-type bands and a large gap above the flat bands in 
$t>1$. 
The origin of the large gap is easily understood 
in a dilute potential limit  $t \rightarrow \infty$. 
In this limit, the potential approaches  one short-range
potential. Its spectrum
consists of a bound state trapped at the potential and a flat band. 
The bound state has the energy $E=V_0$ and the flat band has the energy $E=0$.
In a dense potential limit $t \rightarrow 0$, the potential 
approaches  constant $V_0 /t$. 
Therefore, the asymptotic form of the spectrum in 
$t \rightarrow 0$ and $t \rightarrow \infty$
except for the zero modes
is given by  
\begin{equation}
E(t)\sim V_0 (1+\frac{1}{t}),
\label{asym}
\end{equation}
which satisfies the duality relation (\ref{dual}).

To check the duality, we 
replace $E$ with $Et$ for $t<1$ and replace $t$ with
$2-1/t$ otherwise in Fig.~1. 
Then the figures become  symmetric with respect to
$t=1$ owing to the duality.
The results for the square lattice  are shown in Fig.~2.
The duality is clear in this figure.
The patterns in Figs.~1 and 2 are
very similar to that of the NN and next-nearest-neighbor (NNN)
tight-binding model on the square and the triangular lattice.
However, in contrast with the NN or NNN tight-binding model, a periodicity 
with respect to the flux does not exist in our model. Instead, 
the duality between $t$ and $1/t$ does exist.

As an example of an irregular lattice, we  study a honeycomb lattice 
potential numerically.
A honeycomb lattice  can be regarded as a sum of
two triangular lattices. 
This property causes the following eigenvalue equation at 
$2t={\rm (flux\; per\; hexagon)}/\Phi_0=q/p$:
\begin{equation}
(D_1^\dagger D_1 +D_2^\dagger D_2)\psi=\epsilon \psi,
\end{equation}
where $D_1$ and $D_2$ are defined by
\begin{equation}
D_1=D(p_x,p_y),\quad
D_2=D(p_x-{2\pi\over3p},p_y+{2\pi q\over3}).
\end{equation}
Since both $D_1^\dagger D_1$ and $D_2^\dagger D_2$ are positive definite,
the zero modes must satisfy both $D_1\psi=0$ and $D_2\psi=0$.
If $q-2p>0$, the zero modes exist and the LLL splits into $2p$ subbands
and one flat band.
Otherwise, the LLL splits into $q$ subbands.
This system does not satisfy the duality relation (\ref{dual}).
However, the asymptotic behavior of the one-particle energy spectrum
is also given  by Eq.~(\ref{asym})
from arguments similar to that of  the regular lattice.
The numerical result of the spectrum is shown in Fig.~3.
The spectrum also has a Hofstadter-type structure and a large gap
above the flat bands.

Next we study the possibility of observing
the band structure obtained above.
The gap above the flat bands may be observed in the
magnetoresistance experiment.\cite{ca}
We assume that the antidot potential has a height
$\tilde V_0$ and an area of the base $r_0^2$.
$V_0$ in Eq.~(\ref{peri}) is related to $\tilde V_0$ and $r_0$ as
$a^2V_0=r_0^2\tilde V_0$.
The magnitude of the large gap above the flat bands is of the order
of $V_0$ and the correction of the finite size effect of the antidot to
the energy is estimated as $V_0{\rm O}(r_0/b)$ where $b$ is a lattice
constant of the antidot array.
Therefore the gap above the flat bands should be observed in
$r_0\ll b$.
In the current experiments,\cite{c}
$r_0/b\approx 0.25$, which is enough to observe the gap.
Taking the Landau level mixing into consideration,
the eigenvalue equation for the potential of Eq.~(\ref{peri}) 
has the form
\begin{equation}
\sum_{l'}[E_l\delta_{ll'}+{V_0\over q}D^{(l)\dagger}D^{(l')}]
\psi^{(l')}=\epsilon\psi^{(l)}.
\end{equation}
Because of the $(q-p)$-zero modes of $D^{(l)}$ for $p<q$,
the $l$-th Landau band has a flat band with the energy $E_l$.
Therefore, the flat bands remain flat in the presence
of the Landau level mixing effect in the short-range periodic
potential problem.
The correction of the Landau level mixing to the energy above the
large gap is estimated by the second-order perturbative calculation as
$E_{\rm LL}\approx (V_0)^2/\hbar\omega_c$. From $V_0\gg E_{\rm LL}$,
we obtain the condition $\hbar\omega_c\gg\tilde V_0(r_0/a)^2$ for
the gap to survive.
Using the realistic values\cite{b,ca}
$m=0.07m_e$ and $\tilde V_0=0.3$[meV],
this condition becomes $2\pi\hbar^2/m r_0^2\gg\tilde V_0$.
This is satisfied in the current experiments.
So far, conditions for observation are satisfied.
However, the parameter $t$ in current experiments, {\it e.g.}
$B=5$[T], $b=200$[nm] and $t\approx50$,  
is too large to observe the large gap, the Hofstadter-type bands, and 
the critical behavior near $t=1$.
Thus we hope that the experiment will be made in a finer lattice
of the antidot array.

In summary,
we solved various periodic potential
problems in the LLL using a momentum representation of
the von Neumann lattice.
In a periodic array of short-range potentials, the energy spectrum
has three remarkable structures. 
(i) The spectrum has a Hofstadter-type structure, which is commonly
seen in the periodic potential problems in a magnetic field.
(ii) The duality relation exists for a regular lattice
potential of Eq.~(\ref{peri}).
A honeycomb lattice is an irregular lattice and obeys a duality
relation asymptotically.
(iii) There is a large gap above flat bands\cite{aoki} that comes from
zero modes of the matrix $D$ of Eq.~(\ref{dd}).
These structures are universal in the periodic short-range potential
problem in the LLL.
The conditions for the observation of these structures were obtained.
In addition, the equivalence between two problems of a square lattice of
short-range potentials and a tight-binding model with a inverse flux
was shown.

This work was partially supported by the special Grant-in-Aid
for Promotion of Education and Science in Hokkaido University
provided by the Ministry of Education, Science, Sports and Culture,
the Grant-in-Aid for Scientific Research (07640522), and the
Grant-in-Aid for International Scientific Research (Joint Research
07044048) from
the Ministry of Education, Science, Sports and Culture, Japan.


\vskip10pt
\centerline{
\epsfysize=3.2in\epsffile{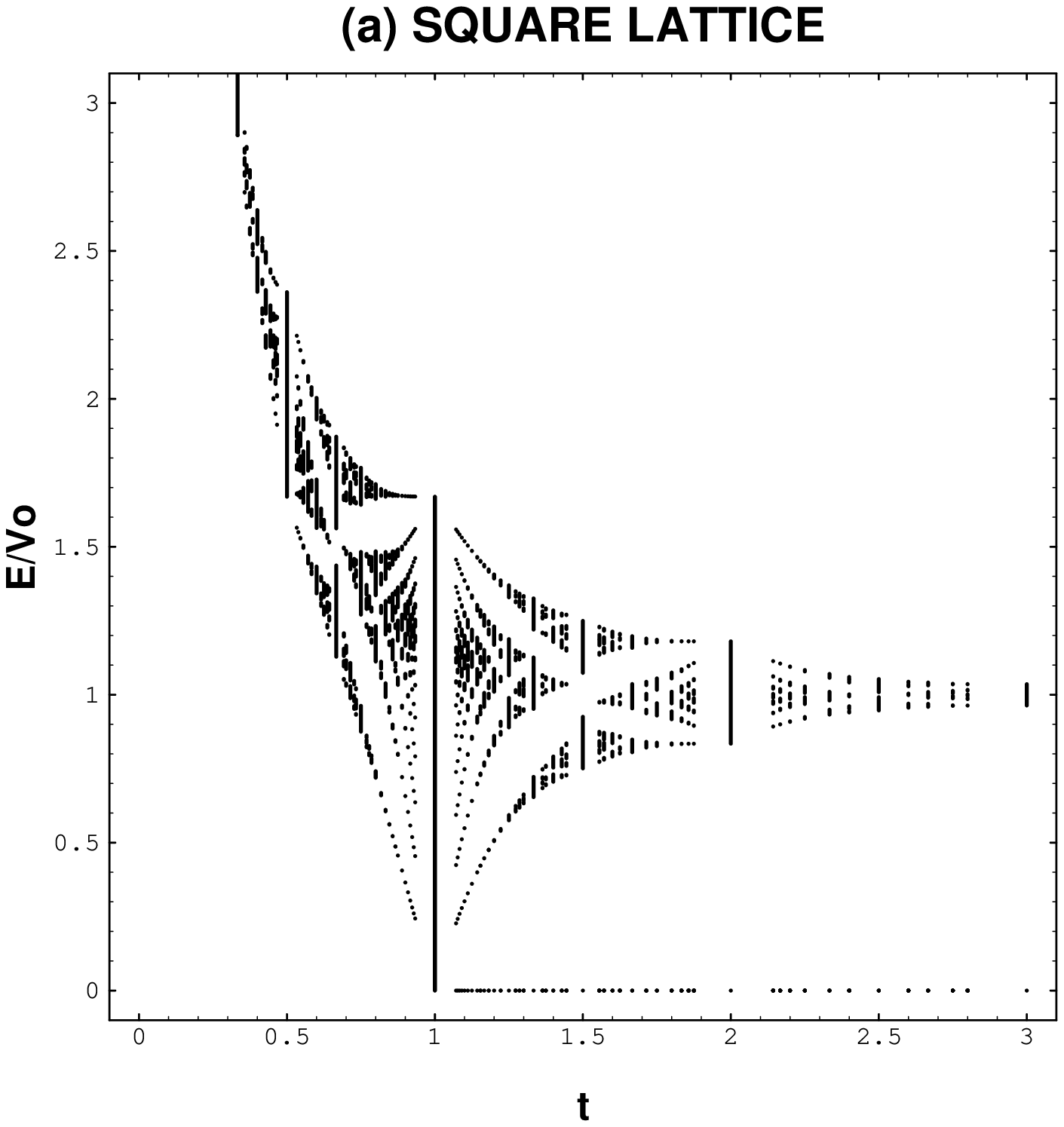}}
\vskip4pt
\centerline{
\epsfysize=3.2in\epsffile{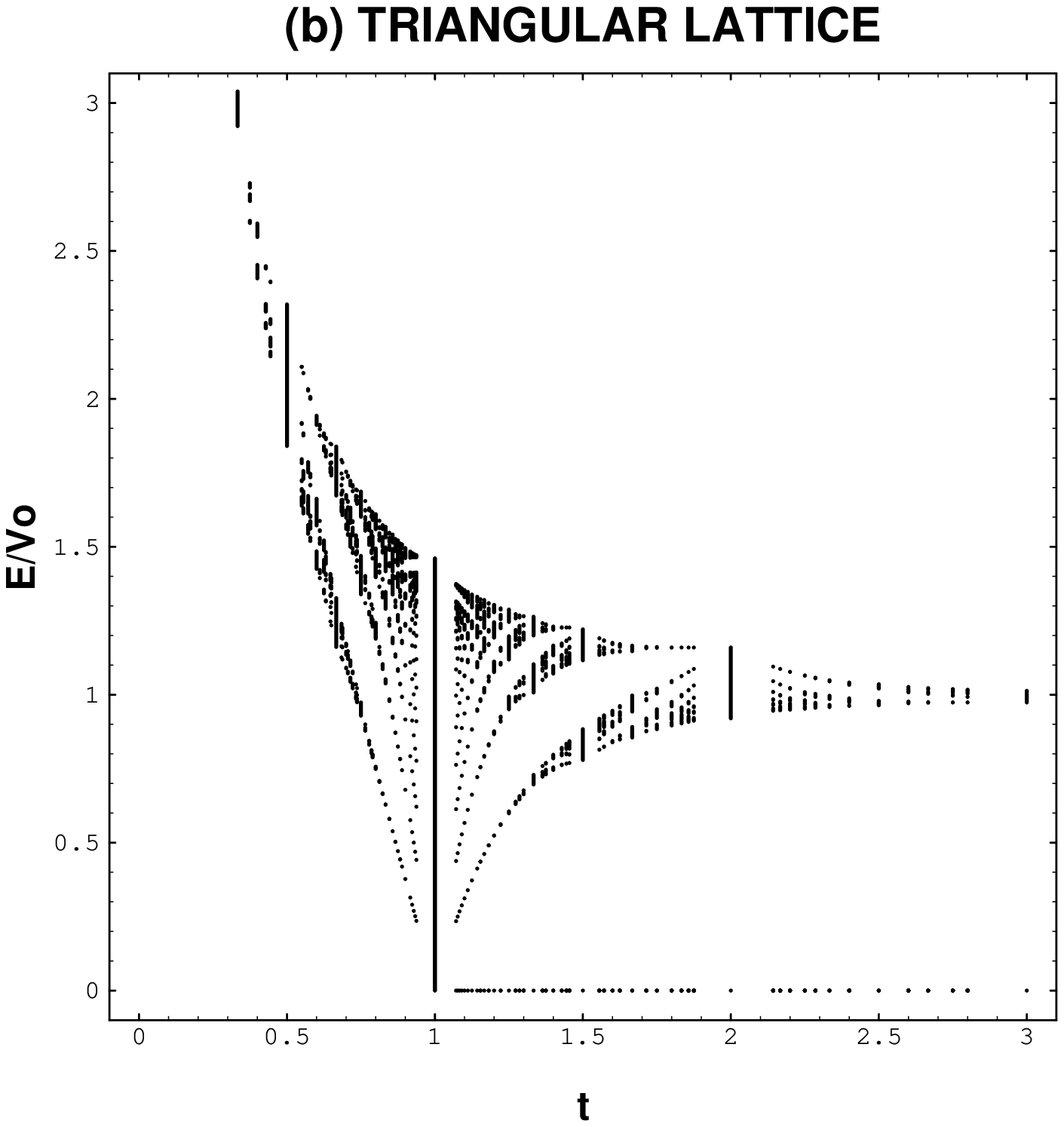}}
\hangindent\parindent{{\small {\bf Figure1}:
Energy spectra of the periodic short-range potential 
for (a) the square lattices and (b) the
triangular lattices. The horizontal variable $t$ is the flux per unit cell 
in the flux unit. $V_0$ is the strength of the potential.
}} 
\vskip10pt
\centerline{\epsfysize=2in\epsffile{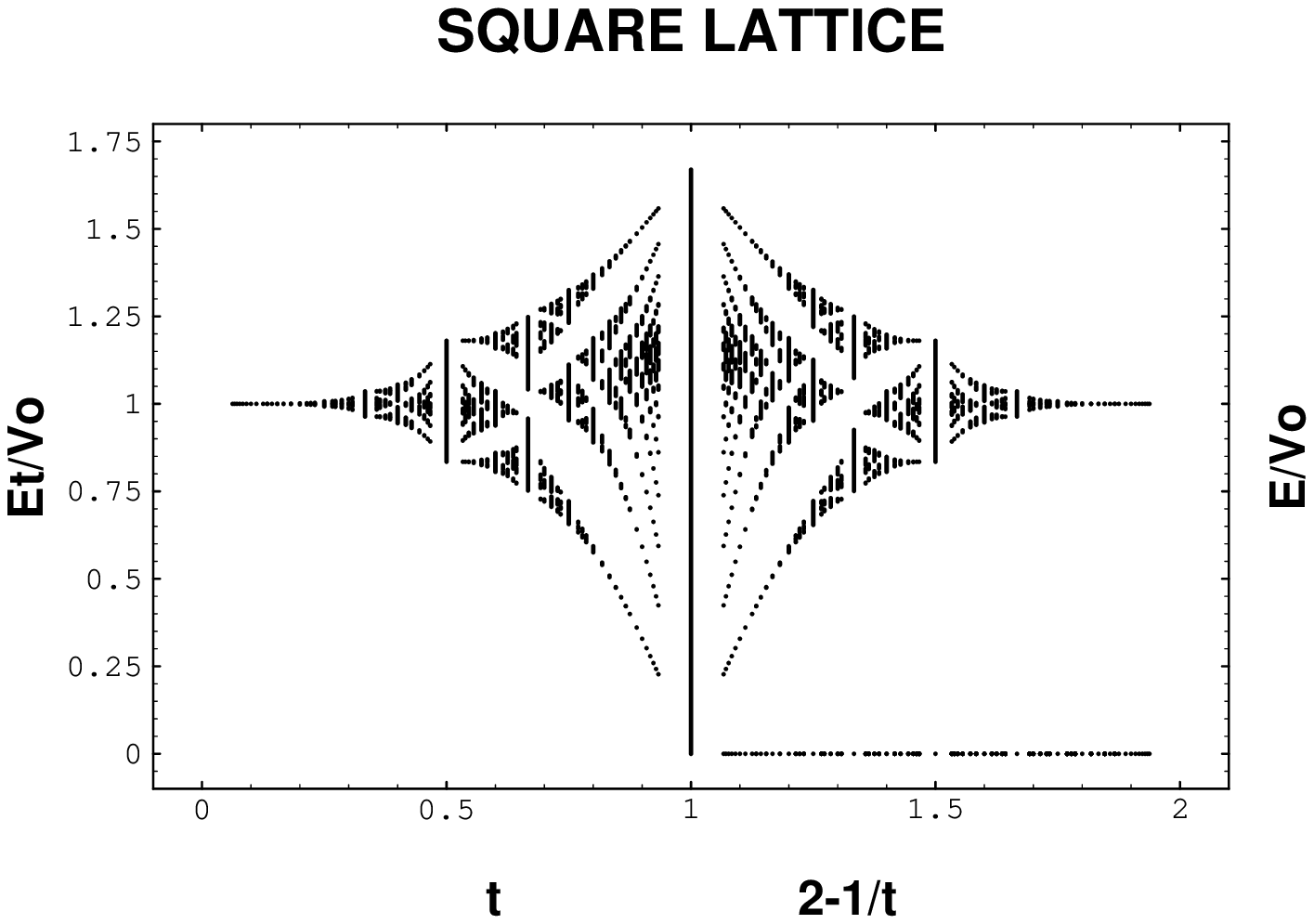}}
\hangindent\parindent{{\small {\bf Figure2}:
Energy spectra of Fig.~1(a) are deformed to show the duality clearly.
The vertical variable is $tE/V_0$ for $t<1$ and $E/V_0$ otherwise.
The horizontal variable is $2-1/t$ for $t>1$ and $t$ otherwise.
As a result, the spectrum becomes symmetric with respect to $t=1$.
}} 
\vskip10pt
\centerline{\epsfysize=3.2in\epsffile{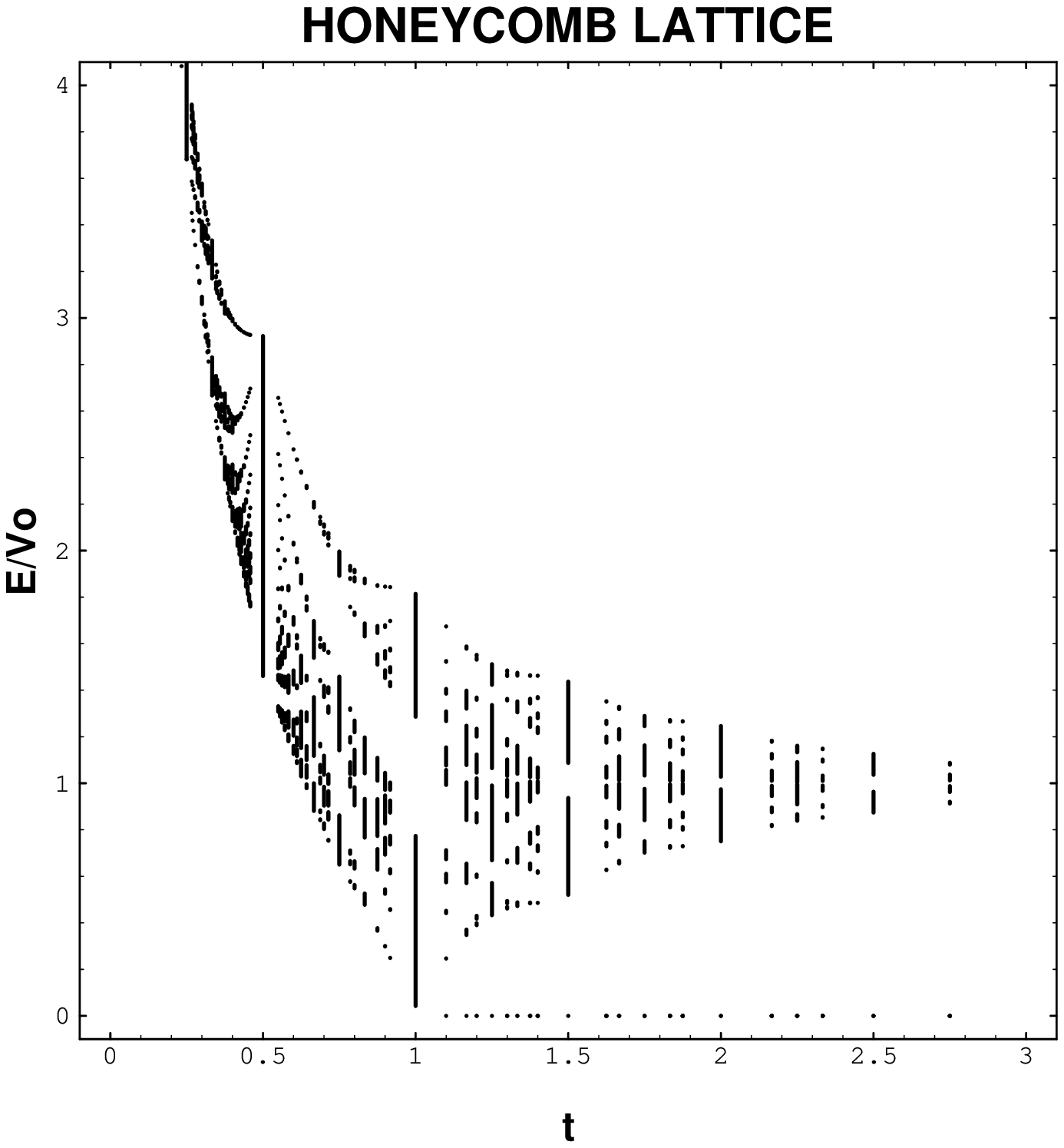}}
\hangindent\parindent{{\small {\bf Figure3}:
Energy spectra for the honeycomb lattices.
The horizontal variable $t$ is the flux per half of the hexagon in the
flux unit. $V_0$ is the strength of the potential.
}} 
\end{document}